\documentclass[12pt,letter,english]{article}

\usepackage[utf8]{inputenc}
\usepackage[T1]{fontenc}
\usepackage{graphicx}
\usepackage{grffile}
\usepackage{longtable}
\usepackage{wrapfig}
\usepackage[normalem]{ulem}
\usepackage{amsmath}
\usepackage{textcomp} 
\usepackage{amssymb}
\usepackage{capt-of}
\usepackage{sidecap}
\usepackage[margin=1in]{geometry}
\usepackage{fancyhdr,xcolor}

\usepackage{color}
\definecolor{colurl}{rgb}{0.7,0.1,0.}
\definecolor{colcite}{rgb}{0.,0.5,0.}
\definecolor{collink}{rgb}{0.4,0.4,0.4}
\usepackage[colorlinks=true,
            breaklinks=true,
            pdfstartview=FitV, 
            linkcolor=collink, 
            citecolor=colcite, 
            urlcolor=colurl, 
            backref=true]{hyperref}
\usepackage{thumbpdf}
\usepackage{morefloats}

\usepackage{tabularx}
\usepackage[T1]{fontenc}
\usepackage{textcomp}
\usepackage{bookman}
\usepackage{wrapfig}
\usepackage{fancyhdr}
\usepackage{graphicx}
\usepackage{amsmath}
\usepackage{amssymb}
\usepackage{ulem}
\usepackage{lscape}
\usepackage{lastpage}
\usepackage{indentfirst}
\usepackage{multicol}
\usepackage{longtable}

\graphicspath{{Figures/}{../Figures/}}
\DeclareGraphicsExtensions{.eps}

\usepackage{color}
\definecolor{grey}{rgb}{0.5,0.5,0.5}

\definecolor{darkgreen}{rgb}{0,0.7,0}

\definecolor{bordeaux}{rgb}{0.5,0.2,0.2}

\definecolor{orange}{rgb}{1,0.5,0}

\newcommand{\eprint}[1]{}

\newcommand{\seppar}{\vspace*{10pt}}

\newcounter{textlistctr}

\newcommand{\squishlist}{
   \begin{list}{$\bullet$}
    { \setlength{\itemsep}{0pt}      \setlength{\parsep}{3pt}
      \setlength{\topsep}{3pt}       \setlength{\partopsep}{0pt}
      \setlength{\leftmargin}{1.5em} \setlength{\labelwidth}{1em}
      \setlength{\labelsep}{0.5em} } }

\newcommand{\squishlisttwo}{
   \begin{list}{$\bullet$}
    { \setlength{\itemsep}{0pt}    \setlength{\parsep}{0pt}
      \setlength{\topsep}{0pt}     \setlength{\partopsep}{0pt}
      \setlength{\leftmargin}{2em} \setlength{\labelwidth}{1.5em}
      \setlength{\labelsep}{0.5em} } }

\newcommand{\squishend}{
    \end{list}  }

\newcounter{obsrefctr}
\newcommand{\kms}{{\,km\,s$^{-1}$}}
\newcommand{\cc}{{\,cm$^{-3}$}}

\newcommand{\1}{{~\sc i}}
\newcommand{\2}{{~\sc ii}}

\newcommand{\4}{{~\sc iv}}

\newcommand{\hmol}{H$_2$}

\def\aj{AJ}%
\def\araa{ARA\&A}%
\def\apj{ApJ}%
\def\apjl{ApJ}%
\def\apjs{ApJS}%
\def\aap{A\&A}%
\def\mnras{MNRAS}%
\def\nat{Nature}%
\usepackage{amssymb}
\usepackage{xcolor}
\definecolor{colred}{RGB}{150,0,0}

\definecolor{colgreen}{RGB}{50,150,50}

\definecolor{colgray}{RGB}{60,60,60}

\usepackage{ragged2e}

\usepackage{times}
\usepackage[numbers]{natbib}
\bibpunct{(}{)}{;}{a}{}{,} 

\begin{document}
\raggedright
\huge
\noindent Astro2020 Science White Paper \linebreak


\bigskip
\bigskip
\noindent\textbf{ISM and CGM in external galaxies}
\normalsize

\bigskip
\bigskip
\noindent\normalsize\textbf{Thematic areas:}  $\boxtimes$ ``Resolved stellar populations and their environments'',  $\boxtimes$ ``Galaxy   Evolution''

\bigskip 
\bigskip
\noindent\normalsize\textbf{Principal Author:}\\
Name: Vianney Lebouteiller$^1$\\
Institution$^1$: Laboratoire AIM, CEA, CNRS, Université Paris-Saclay, Université Paris Diderot, Sorbonne Paris Cité, F-91191 Gif-sur-Yvette, France\\
Email: \textit{vianney.lebouteiller@cea.fr}\\
Phone: +33-1-69-08-49-37 \\

\bigskip
\bigskip 
\noindent\normalsize\textbf{Co-authors:}\\
D. Kunth$^2$, J. Roman-Duval$^3$, P. Richter$^4$, C. Gry$^5$, B. James$^3$, A. Aloisi$^3$ \\
\seppar
$^2$: Institut d'Astrophysique de Paris, Sorbonne Universités, UPMC Univ Paris 6 et CNRS, UMR 7095, 98 bis bd Arago, F-75014 Paris, France\\
$^3$: Space Telescope Science Institute, 3700 San Martin Dr, Baltimore, MD 21218, USA\\
$^4$: Institut f\"ur Physik und Astronomie, Universit\"at Potsdam, Karl-Liebknecht-Str. 24/25, D-14476 Golm, Germany\\
$^5$: Aix Marseille Univ, CNRS, CNES, LAM, Marseille, France\\

\bigskip
\bigskip
\noindent\normalsize\textbf{Abstract:}\\

\justify

The wavelength range $912-2000$\AA\ (hereafter far-UV) provides access to absorption lines of the interstellar medium (ISM), circumgalactic medium (CGM), and intergalactic medium (IGM) in phases spanning a wide range of ionization, density, temperature, and molecular gas fraction. Far-UV space telescopes have enabled detailed studies of the ISM in the Milky Way, paving the way to understand in particular depletion of elements by condensation onto dust grains, molecular gas formation pathways, ionization processes, and the dynamics of shocks. An interesting prospect is to transpose the same level of details to the ISM in external galaxies, in particular in metal-poor galaxies, where the ISM chemical composition, physical conditions, and topology (arrangement of various phases) change dramatically, with significant consequences on the star-formation properties and on the overall galaxy evolution. To circumvent current systematic biases in column density determinations and to examine the ISM enrichment as a function of the environment, we advocate for a versatile far-UV space mission able to observe individual O/B stars or stellar clusters in galaxies up to few $100$s\,Mpc away with a spectral resolution power $R\sim10^{4-5}$. Such requirements would also make it possible to probe multiple quasar lines of sight intersecting the CGM of galaxies at various redshifts, making it possible to reconcile the various physical scales and phases and to comprehend the role of gas exchanges and flows for galaxy evolution.

\pagebreak
\newpage

The far-UV wavelength range provides a unique probe of the interstellar medium (ISM) in external galaxies by enabling the observation of absorption lines arising in clouds either toward individual UV-bright stars, unresolved clusters (e.g., OB associations, H\2\ regions...), or background quasars. Extragalactic lines of sight can also be used to probe the circumgalactic medium (CGM) and intergalactic medium (IGM). The far-UV domain provides access to ISM tracers in various phases and ionization states, including H$^0$, H$_2$, atomic metal species, and molecules, enabling the study of chemical abundances, physical conditions, and molecular gas content in a wide range of galaxy types. Many potential tracers lie shortward of H\1\ Ly$\beta$ (e.g., high-order Lyman series of H\1\ and atomic deuterium $<1025$\AA, Lyman continuum $<912$\AA, strongest \hmol\ lines with weak H\1\ contamination between $1000-1080$\AA).

\section{Metal-poor ISM in nearby galaxies} \label{sec:ism}

\subsection{Solving column density determination biases}\label{sec:biases}

In the far-UV, apart from a few lines of sight toward individual stars in the Magellanic Clouds (e.g., \citealt{Welty2016a,RomanDuval2019a}), the extragalactic ISM has been mostly observed toward stellar clusters and at low spectral resolution ($R\lesssim20\,000$), with inherent biases for the column density determination. First, unresolved absorption lines observed toward a single line of sight may show a low apparent optical depth (i.e., corresponding to the linear regime of the curve of growth) even though some individual velocity components are saturated, leading to the ``hidden'' saturation problem and to the possible underestimate of column densities by factors of a few or more (e.g., \citealt{James2014a}). Second, multiple lines of sight toward stars of different brightness may contribute to the observed spectrum, with each line of sight intersecting multiple interstellar clouds with different properties (metallicity, column density, turbulent velocity, radial velocity). The resulting combination is highly non-linear, especially if some of the individual (i.e., a given line of sight and a given cloud) components are saturated.

While the biases related to saturated components can be mitigated for a single line of sight with a suite of transitions of varying strengths and with a well-behaved distribution of components \citep{Jenkins1986a}, biases related to the multiplicity of lines of sight have been little explored, even in the favorable case of unsaturated components, notably because the spatial distribution of bright stars in the dispersion direction of slit spectrographs complicates even further the line profile and its analysis (e.g., \citealt{Lebouteiller2006a}).  

Hence robust column density determinations in nearby galaxies ideally require a spectral resolution high enough to disentangle $\approx2$\kms\ wide components (typically observed in the Milky Way) and a spatial resolution high enough to resolve individual stars. The corresponding signal-to-noise requirement quickly becomes prohibitive for galaxies further than a few Mpc but satisfactory compromises can be obtained by observing (1) nearby stellar clusters -- such as those observed with HST or FUSE -- with improved spectral resolution $\sim10^5$, (2) distant/faint stellar clusters with spectral resolution $\sim20\,000$, i.e., similar to most current extragalactic ISM spectra, or (3) individual O/B stars with spectral resolution $\sim20\,000$. Another alternative, as proposed by \cite{Kunth1986a}, is to use background QSOs (to be identified with LSST, Euclid...), with similar spectral resolution requirement, provided they are bright enough in rest-frame far-UV where absorption lines in nearby galaxies are probed (e.g., \citealt{Bowen2005a,Kozlowski2013a}).

Overall, an observatory versatile enough to propose high spatial and spectral resolution would solve most of the systematic issues in deriving robust column densities in external galaxies using stellar clusters as background light, thereby nicely complementing studies of Damped Lyman-$\alpha$ systems (DLAs; \citealt{Wolfe2005a}). We review in the following the corresponding science motivations.

\subsection{Chemical abundances}\label{sec:chemab}

Complex lines of sight and limited sensitivity have mostly restricted the study of the extragalactic ISM in the far-UV to resonance lines and chemical abundance determinations, although fine-structure lines can be observed in some nearby galaxies (Sect.\,\ref{sec:sf}). \cite{Kunth1994a} proposed a method to measure neutral gas abundances in blue compact dwarf galaxies (BCDs) using unresolved massive stellar clusters in H\2\ regions as background continuum, thereby providing independent results from abundances traditionally derived using optical emission lines arising in the ionized gas of H\2\ regions. The comparison led to a still ongoing debate, which is well illustrated by several studies of the BCD I\,Zw\,18 ($18$\,Mpc, $\approx2\%$ solar metallicity). Early observations with HST/GHRS showed a discrepancy between the oxygen abundance measured in emission and in absorption, leading to the hypothesis of self-enrichment by the current starburst episode \citep{Kunth1986a,Kunth1994a} but, due to a limited sensitivity, only strong lines were accessible and hidden saturation could not be identified easily \citep{Pettini1995a}. Later studies of the same galaxy with FUSE, which enabled the observation of several metal lines and of hydrogen, highlighted issues regarding the stellar continuum and the selection of weak lines \citep{Aloisi2003a} and showed that only a small discrepancy may exist, if any \citep{Lecavelierdesetangs2004a}. More recently, \cite{Lebouteiller2013a} confirmed, using HST/COS and weak lines such as $\lambda1254$ S\2, that a small discrepancy does exist in I\,Zw\,18. Overall, studies have showed that weak lines (also $\lambda1356$ O\1) may minimize column density determination biases, but at the expense of an in-depth analysis of abundance ratios and of the spatial distribution of metals. 

\begin{SCfigure}[1][t]
  \includegraphics[width=0.5\textwidth]{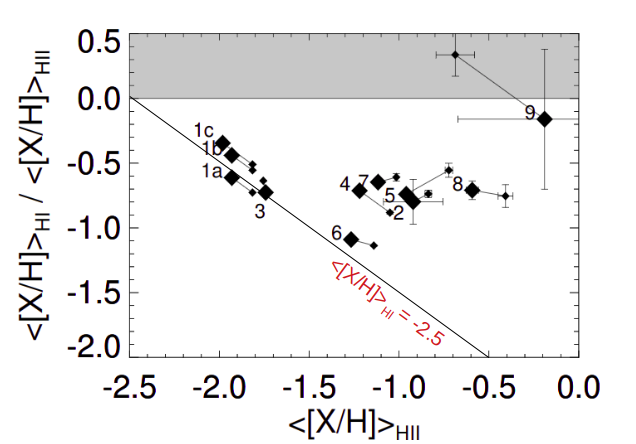}\label{fig}
  \caption{\small Abundance discontinuity between the neutral (HI; observed with FUSE and HST) and ionized (HII) phases in a sample of BCDs \citep{Lebouteiller2009a,Lebouteiller2013a}. Two different methods (diamonds) are compared for each object (numbers). Globally neutral abundances are lower by a factor of a few and the minimum metallicity derived in the neutral phase is around $-2.5$\,dex. } 
\end{SCfigure}

Such discrepancies are important to quantify robustly to understand the distribution of metals and metallicity buildup in galaxies. A sample analysis of neutral gas abundances in BCDs by \cite{Lebouteiller2009a} showed that an overall metallicity floor of $\sim2\%$ solar may exist for galaxies in the nearby Universe which could be linked to the IGM enrichment (Fig.\,\ref{fig}). Metallicity discontinuity between the ionized and neutral phases seem to occur preferentially for moderately metal-poor ($10-50\%$ solar) galaxies, which could be due to dilution by metal-poor/free gas in the halos rather than by self-enrichment in the H\2\ regions (see also \citealt{James2015b}). The presence of quasi pristine gas in the outskirts of galaxies has important implications for the galaxy evolution (e.g., infalling scenarii, dispersal/mixing of elements). For instance, the dust-to-gas mass ratio (D/G) obtained for nearby galaxies shows a steep dependence with metallicity \citep{RemyRuyer2014a} which is at variance with the shallower trend obtained for DLAs (see \citealt{Galliano2018b}). On the one hand, metallicity in nearby galaxies was derived in the ionized gas near young star-forming regions where most of the emitting dust is presumably located. On the other hand, the relative dust abundance in DLAs is derived from the depletion strength of refractory species in the far-UV and the metallicity is derived from lines of sight intersecting the entire galaxy body, also in the far-UV, \textit{including regions that may be quasi-pristine (metal and dust-free)}. Hence the observed D/G vs.\ metallicity slope in DLAs could reflect a dilution factor, with the dust-rich regions having properties in fact similar to the Milky Way value. High spectral resolution is necessary to decompose the velocity profile in metal lines, to infer the corresponding and expected H\1\ absorption, and to compare to the observed one in order to quantify the dilution factor. 

It should be also noted that the determination of elemental abundances in nearby metal-poor galaxies also provide a powerful tool to understand depletion patterns (dust composition) and strengths (dust-to-metal ratio) as a function of metallicity, but such a technique is limited by the small number of metal-poor galaxies with far-UV absorption spectra (see Roman-Duval et al. Astro 2020 white paper). The abundance of deuterium also needs to be explored in the metal-poor ISM, either through D\1/H\1, HD/\hmol, or D/O (the two latter minimizing systematic effects when comparing lines on very different part of the curve of growth; \citealt{Hebrard2002a}), in order to mitigate astration effects (destruction in interiors of stars) and to provide potentially better constraints for Big Bang nucleosynthesis models.

A significant gain in sensitivity is now required to obtain a large sample of metal-poor galaxies (for instance drawn from SDSS; \citealt{Izotov2019a}) while a gain in sensitivity and spatial resolution is required to target individual stars in nearby low-metallicity systems (e.g., Leo\,P, $1.6$\,Mpc, $3\%$ solar; \citealt{McQuinn2015b}) with expected continuum fluxes around $10^{-16}$\,erg\,s$^{-1}$\,cm$^{-2}$\,\AA$^{-1}$. In addition, spectral and spatial resolution are required to solve various biases regarding column density determinations (Sect.\,\ref{sec:biases}) and to determine exact spatial origin of the absorption within the galaxy, which is still unknown. On the other hand, abundances derived from optical emission-lines also suffer from some systematic uncertainties, with a discrepancy observed between abundances derived from collisionally-excited lines and recombination lines (e.g., \citealt{Esteban2002a,Esteban2016a}). This particular discrepancy needs to be explored further by accessing faint recombination lines and abundances in the photosphere of young stars as comparison for various metallicities (e.g., \citealt{Bresolin2016a}).

\subsection{Star-forming gas reservoir}\label{sec:sf}

While most far-UV absorption lines observed in nearby galaxies toward stars or clusters correspond to resonance lines of atomic species, fine-structure atomic transitions and molecular transitions have been detected in a few objects, paving the way to a better understanding of the star-forming gas reservoir. The apparent lack of molecular gas in nearby star-forming low-metallicity galaxies (e.g., \citealt{Taylor1998a,Cormier2014a}) poses fundamental questions regarding the exact role of molecular gas in the star-formation process as compared to the more generally defined cold dense gas (e.g., \citealt{Glover2012a}). While CO is often used to trace H$_2$, it is expected that CO emission is globally weaker in low-metallicity galaxies because of lower C and O abundance and because of selective photodissociation of CO in a dust-poor environment (e.g., \citealt{Wolfire2010a,Schruba2012a}), leading to a potentially large or even dominant reservoir of ``CO-dark'' molecular gas \citep{Grenier2005a}. Accessing both CO and H$_2$ absorption lines would allow a direct measurement of the CO-to-H$_2$ conversion as a function of metallicity and extinction (from translucent to truly molecular), a conversion that is notoriously uncertain. At the same time, other molecules such as OH, CH$^+$, or HD could also be examined as potential tracers of the CO-dark H$_2$ gas.  

Accessing H$_2$ in absorption in molecular clouds is the most direct way to probe molecular gas in low-metallicity environments but it has been limited to translucent clouds ($A_V\sim1-3$), with the lack of diffuse H$_2$ detections in the far-UV in the metal-poor ISM (e.g., \citealt{VidalMadjar2000a}) being explained by enhanced photodissociation and a larger critical surface density for H$_2$ formation \citep{Hoopes2004a,Sternberg2014a}. As observations of molecular clouds in low-metallicity galaxies reach smaller spatial scales, in particular with ALMA, it seems that H$_2$ may exist mostly in dense clumps of size $\lesssim 1$\,pc in such environments (e.g., \citealt{Rubio2015a}, Shi et al.\ in preparation). Such clumps may be identified thanks to near-infrared observations of warm H$_2$ layers (e.g., \citealt{Thuan2004a,Lebouteiller2017a}) but the determination of physical properties (temperature, density, magnetic field, D/G) as a function of the environment (e.g., Milky Way vs.\ low-metallicity galaxies, quiescent vs.\ active star-formation) require the observation of H$_2$ absorption lines in various rotational and vibrational levels. 

Finally, thermal processes can be investigated through the use of absorption lines arising from the fine-structure levels such as C\2*, O\1*, O\1**, Si\2*... Such tracers give valuable information on the ionization degree, temperature, and density of the neutral star-forming gas reservoir and provide indirect constraints on the gas heating mechanisms (photoelectric effect on dust grains, ionization by far-UV or X-ray photons, shocks...) that are independent and complementary to the information provided by far-IR cooling lines. Fine-structure absorption lines have been observed in and around the Milky Way, in Damped Lyman-$\alpha$ systems (shifted to the optical domain), and a few nearby BCDs (e.g., \citealt{Lehner2004a,Wolfe2003a,Howk2005a,Lebouteiller2013a,Lebouteiller2017a}) but the number of Si\2*\ and O\1**\ detections (to measure the gas temperature) remain small due to limited sensitivity. Through fine-structure cooling lines in absorption one can hope to measure the thermal balance in the gas in regions of various extinctions well resolved in space as compared to IR observations, including in low column density infalling filaments/clouds.

\subsection{Nature of compact objects and their influence on the ISM}\label{sec:compact}

The presence of energetic X-ray binaries may influence the ISM properties in galaxies with extremely low D/G and metallicity, with implications for the formation of molecular gas and cold gas and for the star-formation history (see \citealt{Lebouteiller2017a}). The nature of such sources is still debated, though, and the modeling of their properties (including the luminosity in the soft X-rays which deposit their energy in the neutral gas), relies on the absorbing column density toward the X-ray source. An interesting prospect is thus to measure accurately the absorbing column density and ISM metallicity and ionization structure toward X-ray binaries (or toward OB stars in the same region) in nearby galaxies. The identification of compact objects in dwarf galaxies is important as such to probe potential intermediate-mass black holes and to understand whether they participate in the formation of supermassive black holes through coalescence (e.g., \citealt{Mezcua2019a}). Finally, another issue at stake is to understand the relative contribution of Wolf-Rayet stars vs.\ X-ray binaries in the nebular He\2\ emission in low-metallicity star-forming galaxies \citep{Schaerer2019a}.

\section{Gas flows and exchanges in the CGM}\label{sec:cgm}

No model of galaxy formation and evolution is complete without considering gas flows around galaxies. The CGM, in particular, stretches out beyond the virial radii of galaxies and represents a key component of a galaxy’s matter budget that strongly influences its evolution over cosmic timescales, notably through accretion of metal-poor gas and through recycling of enriched galactic material (see e.g., \citealt{Richter2017a,Tumlinson2017a}). The CGM is characterized by its multi-phase nature consisting of cold neutral/molecular gas clumps embedded in diffuse, highly-ionized gas filaments, with wide temperature ($50-5\times10^6$\,K) and density ($10^{-5}-100$\cc) ranges. These gas phases can be probed from low to high redshifts with far-UV absorption and emission lines (including far-UV rest-frame and redshifted extreme-UV transitions).

Past and present far-UV spectrographs (e.g., HST/STIS and HST/COS) have been used to study the CGM along a limited number of quasar lines of sight. However, due to the limitations in sensitivity only one (or at most a few) quasar(s) per galaxy halo are bright enough to spectroscopically investigate the CGM of foreground galaxies, hampering our knowledge of how the CGM functions as a dynamic reservoir for both infalling and outflowing gas. In addition, limited spectral resolution does not enable us to kinematically disentangle the different gas phases, preventing accurate ionization models that are necessary to characterize the physical conditions in the gas and the role of this gas for galactic feedback. 

We now need to determine both the spatial distribution and large-scale kinematics of hydrogen and metal ions in the CGM around low- and high-redshift galaxies, as well as the physical conditions in the gas and its internal density structure. 
Ionization conditions need to be determined for individual phases in order to provide reliable estimates of the total gas mass. Key diagnostic species range from molecular species such as H$_2$ to highly-ionized species such as O\4. At a spectral resolution of $\sim3$\,km\,s$^{-1}$ at $1000$\,\AA, profile fitting of absorption features from metal ions in the CGM directly delivers the temperature of absorbing gas (by resolving thermal line broadening), which then can be used together with the observed ion ratios to self-consistently model the ionization structure of CGM absorbers and their internal gas pressure. On the other hand, the analysis of fine-structure lines such as C\2$^{\star}$ helps to constrain the local cooling rates. Taken together, these measurements would be a crucial step forward to characterize the complex physical conditions that determine the intrinsic properties of CGM clouds and their role in galaxy evolution.

Going back full circle, some of these questions can also be studied by observing the CGM of our own Milky Way toward a large sample of halo stars with known distances, e.g., from GAIA. Are CGM clouds disrupted and incorporated into the halo coronal gas or do they  reach the disk where they can fuel star formation? What are their metallicities and distances? It is worth noting that far-UV observations are well adapted to the measurement of much lower H\1\ ($\gtrsim10^{13}$\,cm$^{-2}$; \citealt{Lehner2006a}) than those accessed with 21\,cm observations.

\pagebreak
\newpage
\setlength{\bibsep}{0.0pt}


\begin{thebibliography}{44}
\expandafter\ifx\csname natexlab\endcsname\relax\def\natexlab#1{#1}\fi

\bibitem[{{Aloisi} {et~al.}(2003){Aloisi}, {Savaglio}, {Heckman}, {Hoopes},
  {Leitherer}, \& {Sembach}}]{Aloisi2003a}
{Aloisi}, A., {Savaglio}, S., {Heckman}, T.~M., {et~al.} 2003, \apj, 595, 760

\bibitem[{{Bowen} {et~al.}(2005){Bowen}, {Jenkins}, {Pettini}, \&
  {Tripp}}]{Bowen2005a}
{Bowen}, D.~V., {Jenkins}, E.~B., {Pettini}, M., \& {Tripp}, T.~M. 2005, \apj,
  635, 880

\bibitem[{{Bresolin} {et~al.}(2016){Bresolin}, {Kudritzki}, {Urbaneja},
  {Gieren}, {Ho}, \& {Pietrzy{\'n}ski}}]{Bresolin2016a}
{Bresolin}, F., {Kudritzki}, R.-P., {Urbaneja}, M.~A., {et~al.} 2016, \apj,
  830, 64

\bibitem[{{Cormier} {et~al.}(2014){Cormier}, {Madden}, {Lebouteiller}, {Hony},
  {Aalto}, {Costagliola}, {Hughes}, {R{\'e}my-Ruyer}, {Abel}, {Bayet},
  {Bigiel}, {Cannon}, {Cumming}, {Galametz}, {Galliano}, {Viti}, \&
  {Wu}}]{Cormier2014a}
{Cormier}, D., {Madden}, S.~C., {Lebouteiller}, V., {et~al.} 2014, \aap, 564,
  A121

\bibitem[{{Esteban}(2002)}]{Esteban2002a}
{Esteban}, C. 2002, in Revista Mexicana de Astronomia y Astrofisica, vol.~27,
  Vol.~12, Revista Mexicana de Astronomia y Astrofisica Conference Series, ed.
  W.~J. {Henney}, J.~{Franco}, \& M.~{Martos}, 56--61

\bibitem[{{Esteban} {et~al.}(2016){Esteban}, {Toribio San Cipriano}, \&
  {Garc{\'{\i}}a-Rojas}}]{Esteban2016a}
{Esteban}, C., {Toribio San Cipriano}, L., \& {Garc{\'{\i}}a-Rojas}, J. 2016,
  ArXiv e-prints [\eprint[arXiv]{1612.03633}]

\bibitem[{{Galliano} {et~al.}(2018){Galliano}, {Galametz}, \&
  {Jones}}]{Galliano2018b}
{Galliano}, F., {Galametz}, M., \& {Jones}, A.~P. 2018, \araa, 56, 673

\bibitem[{{Glover} \& {Clark}(2012)}]{Glover2012a}
{Glover}, S.~C.~O. \& {Clark}, P.~C. 2012, \mnras, 421, 9

\bibitem[{{Grenier} {et~al.}(2005){Grenier}, {Casandjian}, \&
  {Terrier}}]{Grenier2005a}
{Grenier}, I.~A., {Casandjian}, J.-M., \& {Terrier}, R. 2005, Science, 307,
  1292

\bibitem[{{H{\'e}brard} {et~al.}(2002){H{\'e}brard}, {Lemoine}, {Vidal-Madjar},
  {D{\'e}sert}, {Lecavelier des {\'E}tangs}, {Ferlet}, {Wood}, {Linsky},
  {Kruk}, {Chayer}, {Lacour}, {Blair}, {Friedman}, {Moos}, {Sembach},
  {Sonneborn}, {Oegerle}, \& {Jenkins}}]{Hebrard2002a}
{H{\'e}brard}, G., {Lemoine}, M., {Vidal-Madjar}, A., {et~al.} 2002, \apjs,
  140, 103

\bibitem[{{Hoopes} {et~al.}(2004){Hoopes}, {Sembach}, {Heckman}, {Meurer},
  {Aloisi}, {Calzetti}, {Leitherer}, \& {Martin}}]{Hoopes2004a}
{Hoopes}, C.~G., {Sembach}, K.~R., {Heckman}, T.~M., {et~al.} 2004, \apj, 612,
  825

\bibitem[{{Howk} {et~al.}(2005){Howk}, {Wolfe}, \& {Prochaska}}]{Howk2005a}
{Howk}, J.~C., {Wolfe}, A.~M., \& {Prochaska}, J.~X. 2005, \apjl, 622, L81

\bibitem[{{Izotov} {et~al.}(2019){Izotov}, {Guseva}, {Fricke}, \&
  {Henkel}}]{Izotov2019a}
{Izotov}, Y.~I., {Guseva}, N.~G., {Fricke}, K.~J., \& {Henkel}, C. 2019, arXiv
  e-prints [\eprint[arXiv]{1902.01775}]

\bibitem[{{James} \& {Aloisi}(2015)}]{James2015b}
{James}, B.~L. \& {Aloisi}, A. 2015, ArXiv e-prints
  [\eprint[arXiv]{1510.03821}]

\bibitem[{{James} {et~al.}(2014){James}, {Aloisi}, {Heckman}, {Sohn}, \&
  {Wolfe}}]{James2014a}
{James}, B.~L., {Aloisi}, A., {Heckman}, T., {Sohn}, S.~T., \& {Wolfe}, M.~A.
  2014, \apj, 795, 109

\bibitem[{{Jenkins}(1986)}]{Jenkins1986a}
{Jenkins}, E.~B. 1986, \apj, 304, 739

\bibitem[{{Koz{\l}owski} {et~al.}(2013){Koz{\l}owski}, {Onken}, {Kochanek},
  {Udalski}, {Szyma{\'n}ski}, {Kubiak}, {Pietrzy{\'n}ski}, {Soszy{\'n}ski},
  {Wyrzykowski}, {Ulaczyk}, {Poleski}, {Pietrukowicz}, {Skowron}, {OGLE
  Collaboration}, {Meixner}, \& {Bonanos}}]{Kozlowski2013a}
{Koz{\l}owski}, S., {Onken}, C.~A., {Kochanek}, C.~S., {et~al.} 2013, \apj,
  775, 92

\bibitem[{{Kunth} {et~al.}(1994){Kunth}, {Lequeux}, {Sargent}, \&
  {Viallefond}}]{Kunth1994a}
{Kunth}, D., {Lequeux}, J., {Sargent}, W.~L.~W., \& {Viallefond}, F. 1994,
  \aap, 282, 709

\bibitem[{{Kunth} \& {Sargent}(1986)}]{Kunth1986a}
{Kunth}, D. \& {Sargent}, W.~L.~W. 1986, \apj, 300, 496

\bibitem[{{Lebouteiller} {et~al.}(2013){Lebouteiller}, {Heap}, {Hubeny}, \&
  {Kunth}}]{Lebouteiller2013a}
{Lebouteiller}, V., {Heap}, S., {Hubeny}, I., \& {Kunth}, D. 2013, \aap, 553,
  A16

\bibitem[{{Lebouteiller} {et~al.}(2006){Lebouteiller}, {Kunth}, {Lequeux},
  {Aloisi}, {D{\'e}sert}, {H{\'e}brard}, {Lecavelier Des {\'E}tangs}, \&
  {Vidal-Madjar}}]{Lebouteiller2006a}
{Lebouteiller}, V., {Kunth}, D., {Lequeux}, J., {et~al.} 2006, \aap, 459, 161

\bibitem[{{Lebouteiller} {et~al.}(2009){Lebouteiller}, {Kunth}, {Thuan}, \&
  {Désert}}]{Lebouteiller2009a}
{Lebouteiller}, V., {Kunth}, D., {Thuan}, T.~X., \& {Désert}, J.~M. 2009,
  \aap, 494, 915

\bibitem[{{Lebouteiller} {et~al.}(2017){Lebouteiller}, {P{\'e}quignot},
  {Cormier}, {Madden}, {Pakull}, {Kunth}, {Galliano}, {Chevance}, {Heap},
  {Lee}, \& {Polles}}]{Lebouteiller2017a}
{Lebouteiller}, V., {P{\'e}quignot}, D., {Cormier}, D., {et~al.} 2017, \aap,
  602, A45

\bibitem[{{Lecavelier des Etangs} {et~al.}(2004){Lecavelier des Etangs},
  {D{\'e}sert}, {Kunth}, {Vidal-Madjar}, {Callejo}, {Ferlet}, {H{\'e}brard}, \&
  {Lebouteiller}}]{Lecavelierdesetangs2004a}
{Lecavelier des Etangs}, A., {D{\'e}sert}, J.-M., {Kunth}, D., {et~al.} 2004,
  \aap, 413, 131

\bibitem[{{Lehner} {et~al.}(2006){Lehner}, {Savage}, {Wakker}, {Sembach}, \&
  {Tripp}}]{Lehner2006a}
{Lehner}, N., {Savage}, B.~D., {Wakker}, B.~P., {Sembach}, K.~R., \& {Tripp},
  T.~M. 2006, \apjs, 164, 1

\bibitem[{{Lehner} {et~al.}(2004){Lehner}, {Wakker}, \& {Savage}}]{Lehner2004a}
{Lehner}, N., {Wakker}, B.~P., \& {Savage}, B.~D. 2004, \apj, 615, 767

\bibitem[{{McQuinn} {et~al.}(2015){McQuinn}, {Skillman}, {Dolphin}, {Cannon},
  {Salzer}, {Rhode}, {Adams}, {Berg}, {Giovanelli}, \& {Haynes}}]{McQuinn2015b}
{McQuinn}, K.~B.~W., {Skillman}, E.~D., {Dolphin}, A., {et~al.} 2015, \apjl,
  815, L17

\bibitem[{{Mezcua}(2019)}]{Mezcua2019a}
{Mezcua}, M. 2019, Nature Astronomy, 3, 6

\bibitem[{{Pettini} \& {Lipman}(1995)}]{Pettini1995a}
{Pettini}, M. \& {Lipman}, K. 1995, \aap, 297, L63

\bibitem[{{R{\'e}my-Ruyer} {et~al.}(2014){R{\'e}my-Ruyer}, {Madden},
  {Galliano}, {Galametz}, {Takeuchi}, {Asano}, {Zhukovska}, {Lebouteiller},
  {Cormier}, {Jones}, {Bocchio}, {Baes}, {Bendo}, {Boquien}, {Boselli},
  {DeLooze}, {Doublier-Pritchard}, {Hughes}, {Karczewski}, \&
  {Spinoglio}}]{RemyRuyer2014a}
{R{\'e}my-Ruyer}, A., {Madden}, S.~C., {Galliano}, F., {et~al.} 2014, \aap,
  563, A31

\bibitem[{{Richter}(2017)}]{Richter2017a}
{Richter}, P. 2017, in Astrophysics and Space Science Library, Vol. 430, Gas
  Accretion onto Galaxies, ed. A.~{Fox} \& R.~{Dav{\'e}}, 15

\bibitem[{{Roman-Duval} {et~al.}(2019){Roman-Duval}, {Jenkins}, {Williams},
  {Tchernyshyov}, {Gordon}, {Meixner}, {Hagen}, {Peek}, {Sandstrom}, {Werk}, \&
  {Yanchulova Merica-Jones}}]{RomanDuval2019a}
{Roman-Duval}, J., {Jenkins}, E.~B., {Williams}, B., {et~al.} 2019, arXiv
  e-prints [\eprint[arXiv]{1901.06027}]

\bibitem[{{Rubio} {et~al.}(2015){Rubio}, {Elmegreen}, {Hunter}, {Brinks},
  {Cort{\'e}s}, \& {Cigan}}]{Rubio2015a}
{Rubio}, M., {Elmegreen}, B.~G., {Hunter}, D.~A., {et~al.} 2015, \nat, 525, 218

\bibitem[{{Schaerer} {et~al.}(2019){Schaerer}, {Fragos}, \&
  {Izotov}}]{Schaerer2019a}
{Schaerer}, D., {Fragos}, T., \& {Izotov}, Y.~I. 2019, \aap, 622, L10

\bibitem[{{Schruba} {et~al.}(2012){Schruba}, {Leroy}, {Walter}, {Bigiel},
  {Brinks}, {de Blok}, {Kramer}, {Rosolowsky}, {Sandstrom}, {Schuster},
  {Usero}, {Weiss}, \& {Wiesemeyer}}]{Schruba2012a}
{Schruba}, A., {Leroy}, A.~K., {Walter}, F., {et~al.} 2012, \aj, 143, 138

\bibitem[{{Sternberg} {et~al.}(2014){Sternberg}, {Le Petit}, {Roueff}, \& {Le
  Bourlot}}]{Sternberg2014a}
{Sternberg}, A., {Le Petit}, F., {Roueff}, E., \& {Le Bourlot}, J. 2014, \apj,
  790, 10

\bibitem[{{Taylor} {et~al.}(1998){Taylor}, {Kobulnicky}, \&
  {Skillman}}]{Taylor1998a}
{Taylor}, C.~L., {Kobulnicky}, H.~A., \& {Skillman}, E.~D. 1998, \aj, 116, 2746

\bibitem[{{Thuan} {et~al.}(2004){Thuan}, {Hibbard}, \&
  {L{\'e}vrier}}]{Thuan2004a}
{Thuan}, T.~X., {Hibbard}, J.~E., \& {L{\'e}vrier}, F. 2004, \aj, 128, 617

\bibitem[{{Tumlinson} {et~al.}(2017){Tumlinson}, {Peeples}, \&
  {Werk}}]{Tumlinson2017a}
{Tumlinson}, J., {Peeples}, M.~S., \& {Werk}, J.~K. 2017, \araa, 55, 389

\bibitem[{{Vidal-Madjar} {et~al.}(2000){Vidal-Madjar}, {Kunth}, {Lecavelier des
  Etangs}, {Lequeux}, {Andr{\'e}}, {Ben Jaffel}, {Ferlet}, {H{\'e}brard},
  {Howk}, {Kruk}, {Lemoine}, {Moos}, {Roth}, {Sonneborn}, \&
  {York}}]{VidalMadjar2000a}
{Vidal-Madjar}, A., {Kunth}, D., {Lecavelier des Etangs}, A., {et~al.} 2000,
  \apjl, 538, L77

\bibitem[{{Welty} {et~al.}(2016){Welty}, {Lauroesch}, {Wong}, \&
  {York}}]{Welty2016a}
{Welty}, D.~E., {Lauroesch}, J.~T., {Wong}, T., \& {York}, D.~G. 2016, \apj,
  821, 118

\bibitem[{{Wolfe} {et~al.}(2005){Wolfe}, {Gawiser}, \&
  {Prochaska}}]{Wolfe2005a}
{Wolfe}, A.~M., {Gawiser}, E., \& {Prochaska}, J.~X. 2005, \araa, 43, 861

\bibitem[{{Wolfe} {et~al.}(2003){Wolfe}, {Prochaska}, \&
  {Gawiser}}]{Wolfe2003a}
{Wolfe}, A.~M., {Prochaska}, J.~X., \& {Gawiser}, E. 2003, \apj, 593, 215

\bibitem[{{Wolfire} {et~al.}(2010){Wolfire}, {Hollenbach}, \&
  {McKee}}]{Wolfire2010a}
{Wolfire}, M.~G., {Hollenbach}, D., \& {McKee}, C.~F. 2010, \apj, 716, 1191

\end{thebibliography}
\end{document}